Narrowband large amplitude whistler-mode waves in the solar wind and their association with electrons: STEREO waveform capture observations

C. A. Cattell, B. Short, A.W. Breneman, and P.Grul

Running title: Large amplitude oblique solar wind whistlers

Abstract: Large-amplitude (up to 70 mV/m) whistler-mode waves at frequencies of ~0.2 to 0.4 $f_{ce}$ (the electron cyclotron frequency) are frequently observed in the solar wind. The waves are obliquely propagating at angles close to the resonance cone, with significant electric fields parallel to the background magnetic field, enabling strong interactions with solar wind electrons. Very narrowband (sinusoidal waveforms) and less coherent waves (more irregular waveforms) occur. Frequencies and/or propagation angles are distinctly different from whistler-mode waves usually observed in the solar wind, and amplitudes are 1 to 3 orders of magnitude larger. Individual wave packets have durations from a few tenths of a second up to a few seconds. Waves occur most often in association with stream interaction regions (SIRs), and are often 'close-packed.' Sixty-eight percent of the 54 SIRs observed during our study interval had narrowband whistler groups, and seventy-six percent had incoherent groups; thirty-three percent of the nine interplanetary coronal mass ejections (ICMEs) had coherent groups and sixty-seven percent had incoherent groups. In contrast, only nine percent of the thirty-four interplanetary (IP) shocks had whistler groups within 30 minutes of the ramp. Although wave occurrence as a function of the electron temperature anisotropy and parallel beta is constrained by the thresholds for the whistler temperature anisotropy and the firehose instabilities, neither mechanism is consistent with observed wave properties. We show for the first time that comparisons of wave data to thresholds for the electron beam driven instability (beam speed > twice the electron Alfven speed) and to the whistler heat flux fan instability indicate that either might destabilize the narrowband waves. In contrast, the less coherent waves, on average, are associated with zero or near zero heat flux and much higher electron Alfven speeds, without higher energy beams. This suggests that the less coherent waves may be more effective in regulating the electron heat flux, or that the scattering and energization of solar wind electrons by the narrowband waves results in broadening of the waves. The highly oblique propagation and large amplitudes of both the narrowband and less coherent whistlers enable resonant interactions with electrons over a broad energy range, and, unlike parallel whistlers doesn't require that the electrons and waves counter-propagate.

1. Introduction

The importance of whistler-mode waves in the evolution of solar wind electrons has long been a topic of interest. Most theoretical and simulation studies, as well as observational studies (Gurnett and Anderson, 1977; Neubauer et al.,1977; Lin et al., 1998) have concluded that the whistler-mode waves occur at frequencies of much less than the electron cyclotron frequency, and propagate parallel (or anti-parallel) to the solar wind magnetic field. There are only a few observational studies focused on simultaneous measurements of whistler-mode waves and strahl electrons (Gurgiolo et al., 2012; Lacombe et al., 2014; Kajdik et al. 2016), although many examined the consistency of the local or radial dependence of the electrons with theoretical predictions (Feldman et al., 1975; Scime et al.,1994; Graham et al.,2017).

In contrast to these studies, Breneman et al. (2010), using STEREO 3d electric field waveform capture data (time-domain sampler, TDS), discovered very narrowband, large amplitude whistler-mode waves with frequencies of ~0.2 to ~0.4 $f_{ce}$, that were most commonly observed in association with stream interactions regions(SIRs), and at much lower rates with interplanetary (IP) shocks. Assuming the cold plasma whistler-mode dispersion relation, wavelengths were estimated to be ~10 to 25 km (the order of the electron gyroradius), with propagation at angles of 50 to 60 degrees (near the resonance cone), and phase velocities of ~700 km/s to 1200 km/s (comparable to the electron Alfvén speed). These properties indicate that Doppler shifts are not significant. Electric field amplitudes were 10s to up to >100 mV/m, ~1 to 3 orders of magnitude larger than previously observed in the solar wind, with parallel components as much as 30% of the perpendicular component. Because the STEREO instrument does not have search coil data, the phase velocity and wave vector direction (sunward or anti-sunward component) of these waves cannot be directly determined. The wave magnetic fields were estimated to be <~1 nT, comparable to the background magnetic field (though we note that the large parallel to perpendicular electric field ratio indicates that the cold plasma dispersion may not be accurate for obtaining wave parameters). Note that Coroniti et al.(1982), using electric and magnetic field spectral data from ISEE-3, concluded that several wave bursts in a similar frequency band with amplitudes of ~0.1 – 1 mV/m, observed downstream of interplanetary shocks, were consistent with oblique propagation. Given the limitations of the ISEE-3 instrumentation, the properties of the waves, but not the amplitudes, are comparable to the narrowband whistlers reported by Breneman et al.

Most theoretical studies of instability mechanisms for solar wind whistler-mode waves have focused on either temperature anisotropy (Kennel and Petscheck, 1966; Gary and Wang, 1996) or heat flux instabilities (Forslund, 1970; Feldman et al., 1975; Gary, 1978; Gary et al.,1975; Shaaban et al.,2018) They have concluded that only parallel propagating waves at low frequencies (~0.01$f_{ce}$) have significant growth rates. Note many of these studies didn't cover the range of $f_{pe}/f_{ce}$ in our dataset (~35-60). These mechanisms can't explain the frequencies and oblique propagation seen in the STEREO data in Breneman et al. or in this study. Several studies of correlated Langmuir waves and whistler-mode waves (Kennel et al., 1980; Kellogg et al.,1992; Sharma et al., 1992; Ergun et al., 1998)have suggested that either both modes are destabilized by an electron bump-on-tail, or that they may be related by a three-wave decay.

In contrast, the whistler fan instability (Bošková et al., 1992; Krafft and Volokitin, 2003), due to the anomalous cyclotron resonance, generates oblique whistler-mode waves in the appropriate frequency range (~tenths $f_{ce}$). A recent study (Vasko et al., 2019) also examined the whistler fan instability in the context of solar wind core and strahl electrons. Their mechanism could generate obliquely propagating waves of the type reported by Breneman et al. (2010), although Vasko et al. were apparently unfamiliar with the experimental observations of the waves and the associated simulations showing their effectiveness in scattering electrons in the halo and strahl energy range. The instability mechanism is due to strahl electrons, and depends on a number of parameters including the strahl to core density ratio, strahl width, and electron beta. A second proposed mechanism for generating oblique whistler-mode waves is the cyclotron resonance of electron beams with velocities greater than twice the electron Alfven speed ($V_{Ae}$)(Sauer and Sydora, 2010). The waves become more oblique as the beam velocity increases relative to $V_{Ae}$, and waves are damped at high beta.

Several recent studies using ARTEMIS (Tong et al., 2019) or Cluster (Lacombe et al., 2014) data have also shown the existence of whistler-mode waves in this higher frequency band. In contrast to the STEREO results, they find waves that are parallel propagating. Lacombe et al. (2014), using magnetic field spectral data from the Cluster STAFF instrument, found narrow band whistler-mode waves with durations of seconds to several hours. Using events with durations greater that 5 minutes, they concluded that the waves occurred in regions of slow solar wind and low background turbulence, and large electron heat flux, consistent with the whistler heat flux instability, at least for large electron parallel beta. They suggested that the waves could regulate the electron heat flux. A more recent study utilizing ARTEMIS magnetic field waveform data (Tong et al., 2019) also found whistler-mode waves, with occurrence consistent with the heat flux instability. However, they concluded that the amplitudes were too small to have a major role in controlling heat flux. Kajdic et al. (2016) compared electron observations and found that strahl width was broader when narrowband whistlers were observed. Stansby et al. (2016), utilizing electric and magnetic field waveform data from ARTEMIS, determined that large amplitude whistlers had wavelengths dependent on parallel electron beta, and were consistent with the heat flux instability. The waves were propagating parallel to the magnetic field and anti-sunward, and therefore, could not scatter strahl, in disagreement with other studies.

The Breneman et al. study, which found large amplitude obliquely propagating narrowband whistlers, utilized waveform captures with durations of ~0.12s, and thus could not observe the wave packet durations and structure. In this paper, we present a study of STEREO 2.1s waveform captures, enabling us to, for the first time, determine the packet structure of these waves. In Section 2, we present an SIR with examples of the waves and their relationship to electron parameters. In Section 3, we show statistical results on wave properties, association with solar wind structures, and comparison to characteristics of the electrons. In Section 4, we compare to previous studies, theoretical models and discuss the possible importance of these waves for the evolution of solar wind electrons and solar wind structures.

2. Observations of narrowband whistlers

Figure 1 presents an example SIR observed by STEREO-A (note that the SIR extended from 2017 March 15 20:00 UT to 2017 March 17 11:00 UT, but no TDS were obtained until about 2017 March 16 08:00 UT). Panel a plots the magnetic field in RTN coordinates at 8 samples/s(from the IMPACT instrument, Luhmann et al. 2008), panel b shows the proton velocity (black) and density (pink) (from PLASTIC, Galvin et al. 2008), panel c plots the parallel electron beta, $\beta_{e\parallel}$ (black) and the electron temperature anisotropy, $T_{e\perp}/T_{e\parallel}$ (blue), and panel d plots the parallel electron heat flux, $Q_{e\parallel}$. Note that the SWEA data (Sauvaud et al., 2008) that provide the estimates of electron heat flux, beta and temperature anisotropy only cover electrons from ~50 eV to 3 keV. The sample rates for both the PLASTIC data (1 sample/minute) and the SWEA data 2 samples/minute are much slower than the 2.1 s wave observations). The S/WAVES instrument (Bougeret et al., 2008) takes short bursts of 3-d electric field waveforms (TDS) at a commandable set of sample rates and durations with the largest amplitude samples being transmitted. We show data from the ~ 7800 samples/s, 2.1 s mode, enabling study of the wave packet structure of the narrowband whistler-mode waves. Throughout this SIR, 157 TDS were transmitted, of which 109 (at the times indicated by the gold and blue vertical lines) were whistler-mode waves at frequencies of ~0.2 $f_{ce}$, and four were in the ion acoustic frequency range.

In addition to the narrowband whistler-mode waves(NBWM,gold lines), there are waves at comparable frequencies that are broader in frequency with less coherent waveforms (blue lines), which we label as 'incoherent.' Narrowband ('coherent') waves meet the whistler-mode criteria (right-hand polarized, frequency <0.5 $f_{ce}$)and have very sinusoidal waves forms and frequency bandwidth <~10 Hz (see examples in Fig. 1f and g). The bandwidth is defined as two times [maximum frequency at ½ maximum power – minimum frequency over at ½ maximum power). Incoherent waves meet the whistler-mode criteria, but have bandwidths >11 Hz and less coherent waveforms (examples in Fig 1h and i). Of the 109 whistler-waves, 29 met the criteria for coherent, and 80 met the criteria for 'incoherent' The statistics discussed below will examine these two classes of events separately.

A number of characteristics of the oblique whistler waves and their association with SIRs and plasma parameters are illustrated by the example SIR in Figure 1. The waves are often 'close-packed.' In this SIR, the waves occur primarily within the faster solar wind (>450km/s). The waves are not associated with the intervals of largest temperature anisotropy. Coherent waves are observed over a wider range of $\beta_{e\|}$, and often with larger $|Q_{e\|}|$ than the incoherent waves.

3. Statistical results

The database for this study comprises all the 2.1 s TDS obtained by STEREO, including the intermittent intervals from STEREO-A and STEREO-B from June through November, 2011, and continuous period on STEREO-A from March 2017 through January 2018. These TDS were initially run through an automatic wave identification algorithm, based on that of Breneman et al (2010), to identify potential whistler-mode waves, and subsequently examined by eye. Wave characteristics including spacecraft frame frequency, hodograms in magnetic field-aligned coordinates, coherency and power spectra were obtained. Using the cold plasma dispersion relation for whistler-mode waves and the electric field hodograms in minimum variance coordinates (polarization), the wave vector angle to the magnetic field, the phase velocity and the magnetic field perturbation were determined (see Cattell et al., 2008; Breneman et al., 2010 for details). Breneman et al. (2010) showed that Doppler shifts were not significant for these waves. Note that the abnormally large ratio of parallel to perpendicular electric field indicates that the warm plasma dispersion relation would be more accurate. The SIRs, interplanetary coronal mass ejections (ICMEs) and IP shocks identified using the criteria described in Jian et al.(2018, 2019) were obtained from the STEREO Level 3 lists (https://stereo-ssc.nascom.nasa.gov/pub/ins_data/impact/level3/).

Histograms of the wave characteristics are shown in Figure 2, with the left-hand panels referring to NBWM ('coherent') and the right-hand panels to the less coherent('incoherent') whistler-mode waves. Figures 2a and 2b plot the wave frequency normalized to the electron cyclotron frequency ($f/f_{ce}$); 2c and 2d plot the wave amplitude, based on the largest perpendicular component (less than total amplitude due to the large obliquity); and 2e and 2f plot the wave normal angle with respect to the solar wind magnetic field. The average $f/f_{ce}$ is ~0.2 for both; however the width is broader for the incoherent waves, as expected since these was are defined by their broader frequency and less coherent waveforms. The average amplitudes for the incoherent waves is larger (~12 mV m$^{-1}$) than for the coherent (~9 mV m$^{-1}$). Note the maximum observed amplitudes were ~70 mV m$^{-1}$. Both coherent and incoherent waves are highly oblique (average of 64° for coherent and 58° for incoherent), propagating near the resonance cone with the associated significant electric fields parallel to the background magnetic field. The duration of individual wave packets peaks at ~0.1-0.2 s (~10s of wave periods), but extends to ~1.8 s. The average is slightly longer (~0.7 s) for coherent waves compared to incoherent waves (~0.5 s). These results are consistent with Breneman et al. (2010); however, they only studied the coherent waves, and, because they used 0.12 s TDS, they could not determine packet durations.

The occurrence of whistler-mode waves within SIRs and ICMEs is shown in Figure 3, which plots the number of events vs normalized time in the structure with t=0 the start of the structure and t=1 the end of the structure for SIRs and ICMEs. Both coherent and less coherent waves are seen throughout SIRs and ICMEs, with somewhat more in the first half to two-thirds of an event. The occurrence rate of coherent waves with respect to incoherent waves is higher in SIRs. We can also look at the percentage of SIRs and of ICMEs that have these high frequency whistler-mode waves, summarized in Table 1. During the time period when STEREO-A was continuously in the 2.1 s TDS mode, there were 54 SIRs and 9 ICMEs; 68% of the SIRs had coherent whistler groups (defined as 2 or more waves separated by no more than 10 minutes, and an average wave density greater than or equal to one wave per minute). Of the nine ICMEs, 33% had coherent groups. The coherent whistlers are more common in SIRs that ICMEs (consistent with Breneman et al., 2010). However, when we include both coherent and incoherent whistlers, we find that 76% of SIRs, and 67% of ICMEs had whistler groups. For the 34 interplanetary (IP) shocks, three had whistler wave groups within 30 minutes of the shock ramp for a 9% occurrence rate (not shown). Most were seen in the ramp or within ~6 minutes in the downstream region. The low rate is consistent with Breneman et al., and the specific association with the ramp is consistent with Cohen et al. (2019). Note that only 1 out of 54 SIRs and 1 out of 9 ICMEs did not contain at least one whistler wave (respectively 98% and 89% had whistlers); however, 15 of 34 IP shocks did not have whistlers (~55% had whistlers).

The relationships of the waves to solar wind plasma parameters including electron beta, temperature anisotropy, and heat flux, and solar wind speed provide important clues to the instability mechanisms, and the effect of the waves on solar wind electrons. For some plots, the proton density (at an even lower cadence of once/minute than the electron measurements) was used as a better estimate for total solar wind density. The magnetic field data are at 8 samples/s. Figure 4a and 4b shows the dependence of wave occurrence on electron temperature anisotropy ($T_{e\perp}/T_{e\parallel}$) versus parallel electron beta ($\beta_{e\parallel}$). The upper line is the whistler temperature anisotropy threshold, $T_{e\perp}/T_{e\parallel} = 1+0.27/\beta_{e\parallel}^{0.57}$ and the lower line plot is an arbitrary firehose instability (both from Lacombe et al. 2014, based on Gary et al., 2006). The contours indicate the density of observed whistler-mode waves. There is no clear distinction between the two wave types. Although wave occurrence is constrained by the thresholds for the whistler temperature anisotropy and the firehose instabilities, neither mechanism is consistent with observed wave properties. No relationship between the amplitude of waves and the temperature anisotropy ($T_{e\perp}/T_{e\parallel}$) was found for either wave type. Most waves occurred when $T_{e\perp}/T_{e\parallel} < 1$ (see Table 2).

Earlier studies (Lacombe et al., 2014) found that coherent parallel propagating whistler-mode waves occurred within the quiet, slow solar wind. For the oblique waves, the occurrence distribution is centered around ~450 km

s$^{-1}$ (Table 2),but significant numbers of wave packets occur in association with solar wind speeds above 600 km s$^{-1}$.

The dependence of wave occurrence on heat flux and parallel electron beta is plotted in Figure 4c for coherent and Figure 4d for incoherent waves. The threshold for the heat flux instabilitycfrom Gary et al. (1999) is over-plotted in green ($0.5/\beta_{e\parallel}0.8$). Only 11% of the coherent waves and 6% of the incoherent waves are above this threshold. The orange line is the upper limit from Lacombe et al.(2014) for their parallel-propagating whistler-mode waves. It is interesting to note that only 2 of the oblique waves in our study are above this orange line. The dependence on the parallel heat flux, $Q_{e\parallel}$, and the total heat flux are distinctly different for the coherent and incoherent waves, in contrast to the dependence on temperature anisotropy. The number of coherent wave events has a broad peak around 0.006-0.01 ergs/cm$^2$ s, and extends to higher $\beta_{e\parallel}$; however the occurrence of less coherent waves (4d) peaked near 0 ergs/cm$^2$ s, and primarily $\beta_{e\parallel}$ <~0.5. Table 2 summarizes the average values of total heat flux and total beta for both wave types. The very clear difference suggests several possible conclusions: (1) the less coherent waves are more effective at regulating the electron heat flux; (2) the less coherent waves are associated with a different instability mechanism; or (3) there are nonlinear processes that relate the coherent and less coherent wave types. Note there is no clear dependence of wave amplitude on heat flux.

The bottom panel of Figure 5 (e) plots, for the event shown in Figure 1, the energy corresponding to twice the electron Alfven speed, E$_{Ae}$, (black) and dotted lines corresponding to the center energy of SWEA channels (from ~400 eV to ~1100 eV). The locations of coherent waves are over-plotted in gold and of incoherent whistler-mode waves in blue. Most of the narrowband (coherent) waves occur when E$_{Ae}$ is lower than for most incoherent waves. The next set of panels plot pitch angle distributions for SWEA energy bands centered on ~400 eV (5a), 650 eV (5c), and 1057 eV (5b). All the coherent waves are associated with strahl up to center energy of 1067 eV, above the highest E$_{Ae}$ of ~650 eV. In contrast, incoherent waves are seen in regions where there is no strahl or the strahl is at energies below E$_{Ae}$, which ranges from ~1057 eV to ~1820 eV. The average of the 4 STE detectors, plotted in 5a, shows that there were not significant electron fluxes at higher energies. Panel 5a also plots the electron (blue) and ion (red) temperatures. The whistler waves occur in association with higher ion temperatures, possibly indicating that they heat the ions or that the conditions required for wave instability are correlated with higher ion temperatures.

The statistical dependence of wave occurrence on $E_{Ae}$ is shown in Figure 6 (left panel for coherent and right panel for incoherent whistlers. As in the specific example shown in Figure 5, coherent waves are more likely to occur when $E_{Ae}$ is lower than for incoherent waves. The average $E_{Ae}$ is ~175 eV for coherent waves, and ~474 eV for incoherent waves. This suggests that lower energy beams are needed to produce the narrowband waves. Since strahl in this energy range is more commonly observed, the threshold for the beam mechanism will more often be met than for the more incoherent waves. Sauer and Sydora (2010) also found that the waves became more oblique as the beam energy increased with respect to $E_{Ae}$. There is a tendency for the wave angle to increase as average $E_{Ae}$ decreases. Both the dependence of wave occurrence and of wave angle on $E_{Ae}$ are consistent with this beam driven mechanism. A more detailed comparison would require determining the peak energy of the electrons observed at the time of the waves, and whether there is evidence for beam distributions, which we can't include in this study due to limitations imposed by the STEREO electron instruments.

The whistler fan instability has also been shown to destabilize very oblique whistler-mode waves. The dependence of wave occurrence on a proxy for the ratio of the electron heat flux normalized by $Q_0 = 1.5 N_0 m_e v_c^3$ (where $N_0$ is the core density approximated by the proton density and $v_c$ is the core speed, using the value from Wilson et al.,2019) versus total electron beta is shown in Figure 7 left panel (coherent) and right panel (incoherent). The colored lines are the linear instability thresholds from equation 5 (Vasko et al., 2019) for the parameters in their Table 1 and plotted in their Figure 4b. The total number of coherent whistler wave packets is 1810, and there were 721 above the lowest threshold (green line), corresponding to ~40%, and 184 (~10%) above the highest threshold plotted (orange line). Of the 2704 incoherent whistler packets, 497 (~18%) were above the lowest threshold, and only 39 (~1%) above the highest threshold. Given the assumptions needed to obtain values for the normalized heat flux and total electron beta, the fact that the such a large fraction of the coherent waves are above the threshold is very striking. The good correlation indicates that the fan instability is consistent with the STEREO observations, at least for the coherent waves.

4.Discussion and conclusions

The results of the first statistical study of the STEREO 2.1s waveform capture 3d electric field data have shown that both very narrowband

('coherent', bandwidth <10Hz) whistler mode waves (originally identified by Breneman et al, 2010) and less coherent (bandwidth >11 Hz) are very common in SIRs, less common in ICMEs, and rare in association with IP shocks. Both wave types have average frequencies of ~0.2 $f_{ce}$, and are quasi-electrostatic, propagating near the resonance cone (~65° for 'coherent' and 60° for 'incoherent'). Mean amplitudes are slightly larger for the less coherent wave packets (~12 mV m$^{-1}$) than for the narrowband waves (~9 mV m$^{-1}$); peak values for both reach ~70 mV m$^{-1}$. Although STEREO does not have a search coil magnetometer, the magnetic perturbation can be estimated using the cold plasma dispersion relation. Values are often the order of the background solar wind magnetic field, although the highly oblique propagation and significant electric field parallel to the solar wind field indicates that the warm dispersion relation would be more accurate.

The fact that the waves described herein are oblique could be due to propagation effects since whistlers propagating in an inhomogeneous medium can refract towards the resonance cone; however, the fact that the oblique waves are significantly larger amplitude than found in studies of parallel propagating waves in the solar wind makes this a very unlikely explanation. The distribution of wave angles, and the fact that waves are often observed in regions without strong inhomogeneity also contradict this explanation.

Earlier observational studies based on spectral data (Gurnett and Anderson, 1977; Neubauer et al.,1977; Coroniti et al., 1982: Lin et al., 1998) found generally low amplitudes (1 to 3 orders of magnitude smaller than we show), likely due to the long time averages and limited frequency resolution). We showed packet durations are usually <1 second, shorter than the time resolution of older wave instruments. The waves are also very narrowband, which would also result in underestimates of wave amplitudes from spectral data.

The dependence of wave occurrence and amplitudes on $T_{e\perp}/T_{e\parallel}$, $Q_{e\parallel}$, $\beta_{e\parallel}$, $Q_e$, $\beta_e$ and $E_{Ae}$ was compared to theories of instability mechanisms and to previous results. No relationship was found between the amplitude of waves and the temperature anisotropy ($T_{e\perp}/T_{e\parallel}$) for either wave type. Most waves occurred when $T_{e\perp}/T_{e\parallel} < 1$. Although wave occurrence as a function of the electron temperature anisotropy and parallel beta is constrained by the thresholds for the whistler temperature anisotropy and the firehose instabilities, neither mechanism is consistent with observed wave properties.

By comparing our observations to thresholds for the electron beam driven instability proposed by Sauer and Sydora (2010) and the whistler heat flux fan instability (Vasko et al., 2019), we show, for the first time, that either might destabilize the narrowband waves. The comparison to the fan instability (shown in Figure 6) is particularly compelling, given the limitations on the electron measurements. In contrast, the less coherent waves, on average, are associated with zero or near zero heat flux and much higher electron Alfven speeds, without higher energy beams. This suggests that the less coherent waves may better regulate the electron heat flux than the coherent waves, and that there is some evolutionary process connecting the two wave types. Given the similarity in the wave properties, it is less likely that the instability mechanisms for the two are different.

There have been several studies focused on higher frequency(~0.02 – ~0.5 $f_{ce}$) whistler-mode waves in the solar wind, using data from Cluster, ARTEMIS or Wind. The Cluster and ARTEMIS statistical studies all found parallel-propagating whistlers consistent with the heat flux instability, and, therefore, did not focus on the wave modes described in our study. Lacombe et al (2014), in a study of 10 minute intervals in the free solar wind utilizing primarily 4 s spectral data from the Cluster STAFF instrument, found narrowband parallel propagating whistler-mode waves, lasting for intervals of >~5 minutes. For large parallel electron beta, wave occurrence fell along the heat flux instability threshold. The waves occurred in the quiet slow(<500 km/s) solar wind, in association with high electron heat flux. Also using Cluster data, Kadjic et al. (2016) showed that whistler-mode waves (at ~0.1 $f_{ce}$ and propagating within 20 degrees of the magnetic field) observed in pristine, and primarily slow, solar wind resulted in significant broadening of strahl electrons, when compared to intervals without waves. The effect was energy dependent and largest for energies of ~50 times the electron thermal energy.

Stansby et al.(2016)analyzed several whistler electric and magnetic field waveforms obtained in the ARTEMIS burst mode. The wave electric field amplitudes (<~0.2 mV m$^{-1}$) were an order of magnitude below our amplitude threshold (~3 mV m$^{-1}$) Use of both electric and magnetic fields enabled determination of the wave propagation direction, which was magnetic-field aligned and anti-sunward; thus the waves they observed can't interact resonantly with the anti-sunward strahl. Wave properties were consistent with the whistler cold dispersion relation, but the dependence on the parallel electron beta was consistent with the warm dispersion relation. Tong et al.(2019b), utilizing magnetic field spectral data (once every 8s)from ARTEMIS, also concluded that the whistlers were parallel propagating, primarily observed in slow solar wind, and small amplitude ($\frac{\delta B_w}{B_0} < .02$). Occurrence was strongly dependent on the electron temperature anisotropy. The beta dependence of f/$f_{ce}$ was found to be consistent with the heat flux instability, and $T_{e\perp}/T_{e\|} > 1$ was also required. This is not consistent with our results that show most events occur when $T_{e\perp}/T_{e\|} < 1$). Simulations of the heat flux instability (Kuzichev et al, 2019) found that the nonlinear development was consistent with ARTEMIS and Cluster observations.

In contrast to the above studies, several event studies utilizing Wind (TDS) waveform data reached different conclusions. Ergun et al. (1998), in a study of solar Type III radio bursts, showed that the observed electron distributions peaked at ~9 keV were marginally unstable to both oblique, quasi-electrostatic whistlers and to Langmuir waves. They suggested that these oblique waves might play an important role in the evolution of flare-accelerated electrons and other solar wind electrons. This idea is consistent with our results; however, we do not have pitch angle distributions extending to such high energies, and most wave packets were not associated with significant fluxes at energies >2 keV (energy range of the STE detector). In another study of Type III radio bursts, Moullard et al. (1998) observed whistler-mode waves, in this case parallel-propagating, and also suggested either a beam-driven or wave decay mechanism. Associated with a magnetic cloud, Moullard et al. (2001) observed parallel propagating whistlers (~0.3-0.5 $f_{ce}$). The observed waves were small amplitude (~0.1-0.4 mV m$^{-1}$, and dB~0.2nT), and occurred when there was an enhanced loss-cone distribution in hot electrons. They concluded that both the whistlers and the simultaneously observed Langmuir wave packets could be excited by the loss cone distribution. They also speculate that the whistlers might be associated with decay of the Langmuir waves, while stating that this would likely require oblique whistlers. Although this mechanism may operate at times, it cannot explain the whistlers we observed, which were not associated with Langmuir waves.

To summarize, large amplitude, highly oblique whistler-mode waves are commonly observed in the solar wind. The waves are observed over a wide range of solar wind speeds, up to 700 km/s. We have shown, for the first time, that these waves are consistent with both the heat flux fan instability and the electron beam instability. The fact that the less coherent waves occur with zero or low heat flux suggests that they may be more effective in regulating the electron heat flux, or that the scattering and energization of solar wind electrons by the narrowband waves results in broadening of the waves. The highly oblique propagation and large amplitudes of both the narrowband and less coherent whistlers enable resonant interactions with electrons over a broad energy range, and, unlike parallel whistlers, do not require that the electrons and waves counter-propagate.

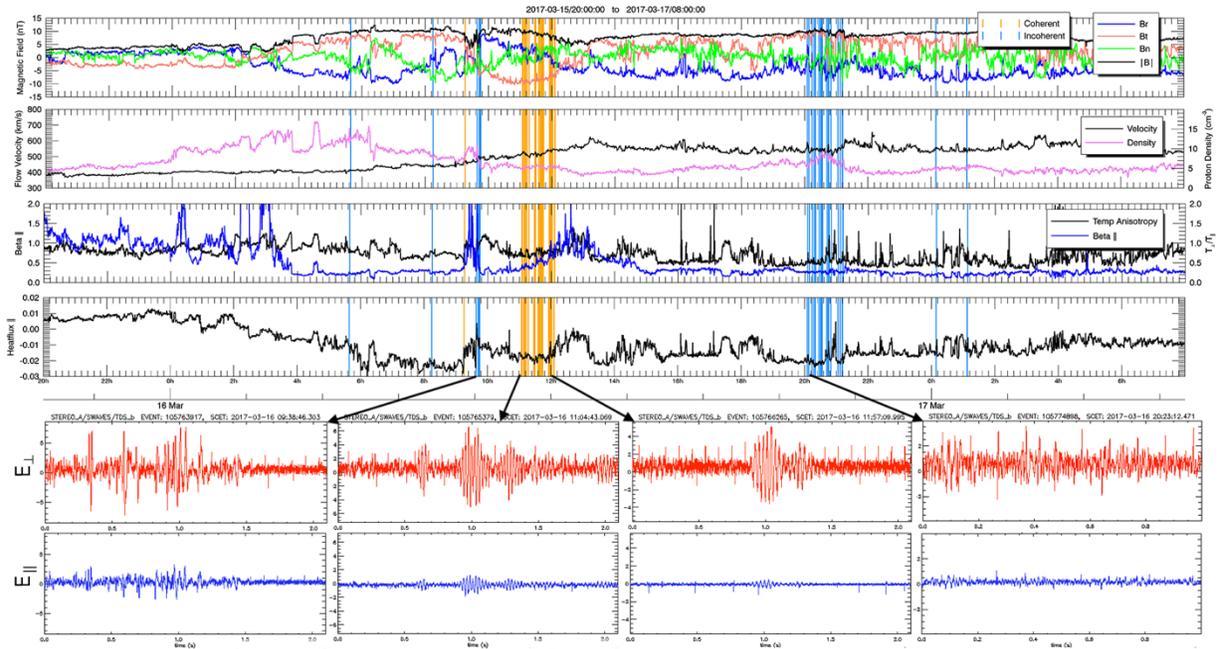

Figure 1. Example of SIR on 2017 March 16. From top to bottom: Three RTN components and magnitude of the magnetic field; proton density (pink) and speed (black); electron temperature anisotropy (blue) and parallel beta (black); parallel electron heat flux. The locations of coherent whistler wave packets are over-plotted in gold; incoherent whistler packets are over-plotted in blue. The bottom set of panels show examples of one perpendicular component of the electric field whistler waveform in red and the parallel component in blue for two incoherent and two coherent waveforms taken at the times indicated by the arrows.

b

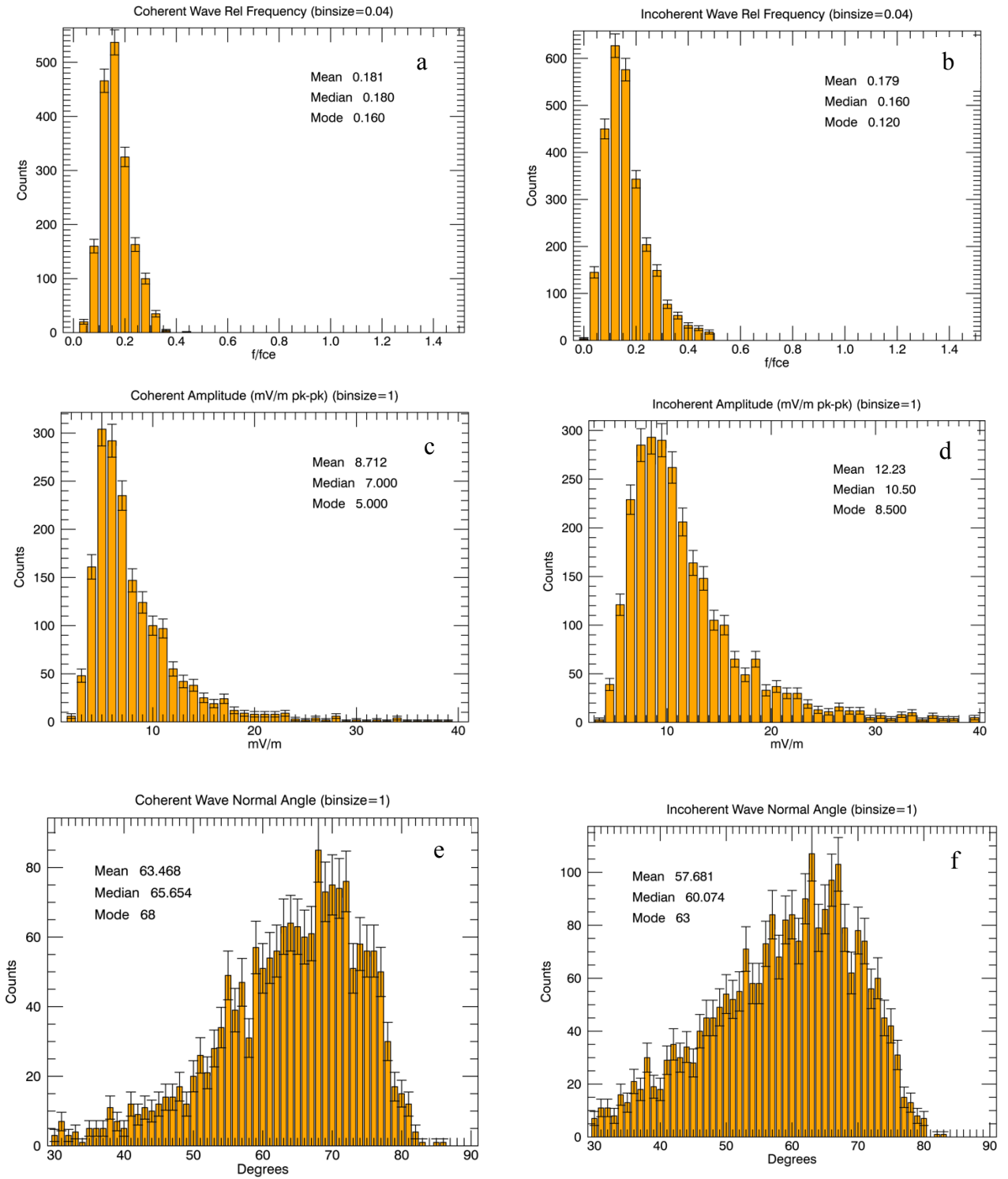

Figure 2. Histograms of wave properties for coherent (left panels) and incoherent (right panels) whistler-mode waves. (a) and (b) $f/f_{ce}$; (c) and (d) amplitude; (e) and (f) wave angle with respect to B.

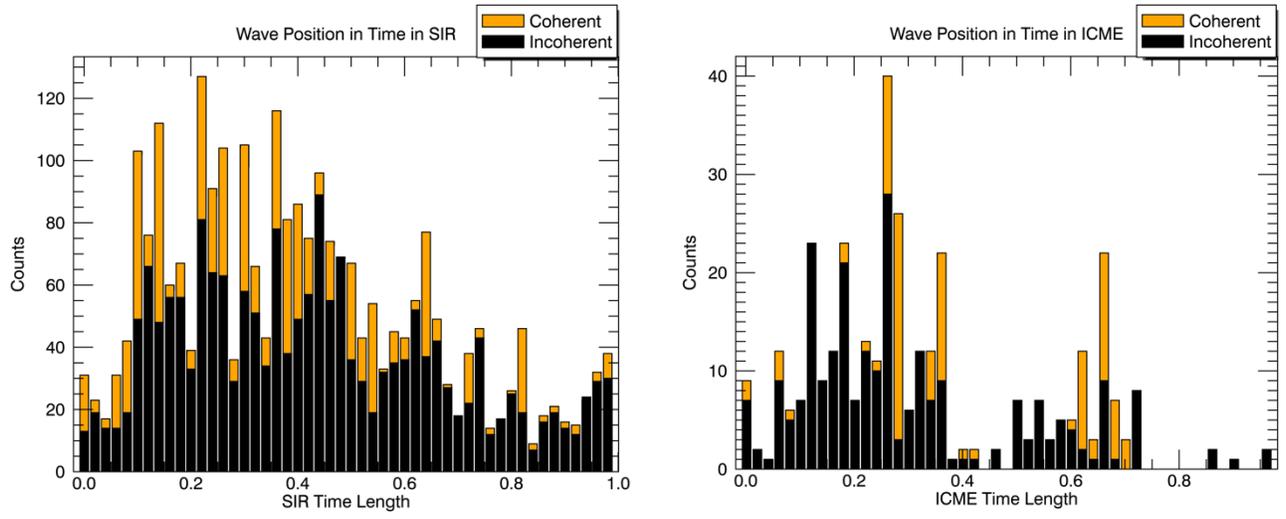

Figure 3. Superposed epoch analysis of the occurrence of whistler waves in stream interactions regions (left) and interplanetary coronal mass ejections(right). Start time of each structure =0; end time=1.

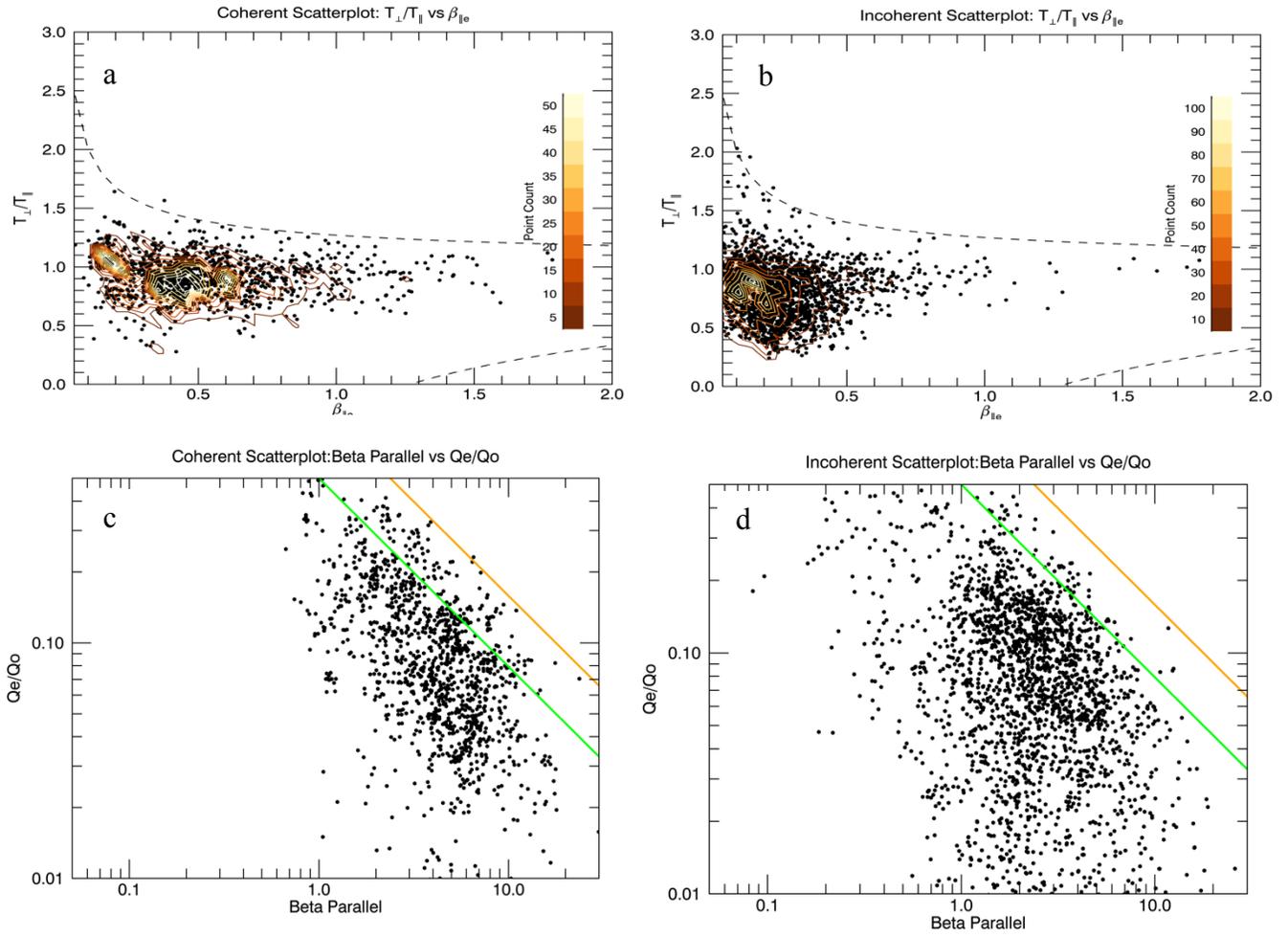

Figure 4. Wave occurrence for electron temperature anisotropy vs parallel beta for (a) coherent and (b) incoherent wave packets. Upper dotted line is the the whistler temperature anisotropy threshold, $T_{e\perp}/T_{e\parallel} = 1+0.27/\beta_{e\parallel}^{0.57}$ and the lower line plot is an arbitrary firehose instability (both from Lacombe et al. 2014, based on Gary et al., 2006). Wave occurrence for normalized electron heat flux vs parallel beta for (c) coherent and (d) incoherent wave packets. Lower green line is the heat flux instability threshold from Gary et al. (1999) and the upper orange line is the upper bound found by Lacombe et al.(2014)

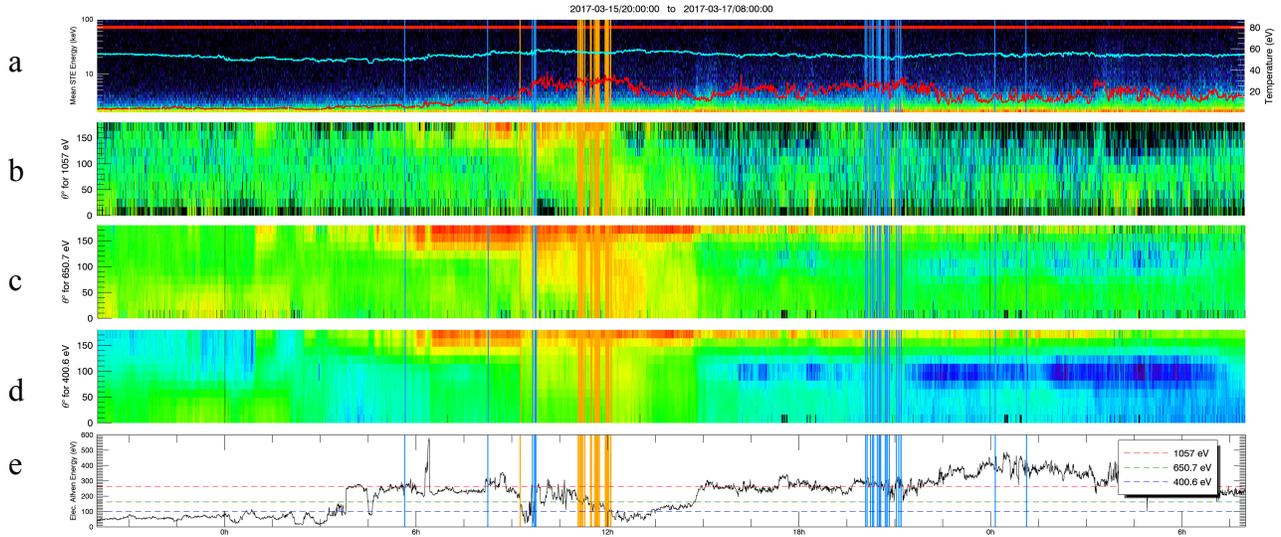

Figure 5. Electron data and Alfven energy for SIR in Figure 1. (a) Energy spectrogram of the average of the 4 STE detectors; ion temperature over-plotted in red; electron temperature in green. Note scale for temperatures (on right) is linear and different from electron energy (on left). (b) pitch angle distribution for for SWEA energy bands centered on 1057 eV (b), 650 eV (c), and ~400 eV (d). (e) the energy corresponding to twice the electron Alfven speed, $E_{Ae}$, (black) and dotted lines corresponding to the center energy of SWEA channels (from ~400 eV to ~1100 eV). The locations of coherent whistler wave packets are over-plotted in gold; incoherent whistler packets are over-plotted in blue.

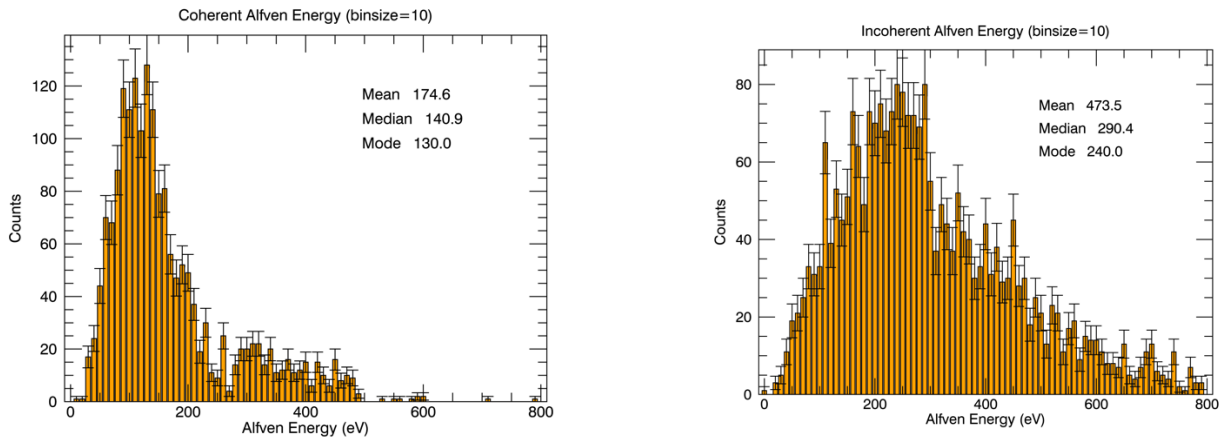

Figure 6. Histogram of wave occurrence vs Alfven energy for coherent (left) and incoherent (right) wave packets.

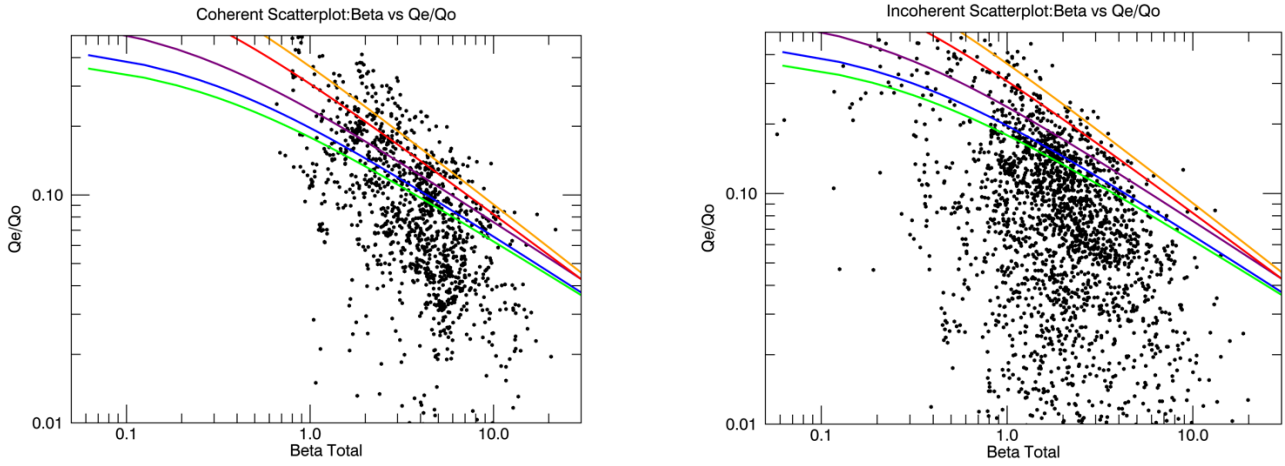

Figure 7. Occurrence of coherent (left) and incoherent (right) whistler waves: normalized electron heat flux vs beta. The colored lines are the linear instability thresholds from equation 5 (Vasko et al., 2019) for the parameters in their Table 1

TABLES

|  | number | Coh. wave group | Any wave group | ≥1 whistler |
|---|---|---|---|---|
| SIRs | 54 | 68% | 76% | 98% |
| ICMEs | 9 | 33% | 67% | 89% |
| IP shocks | 34 |  | 9% | 55% |

Table 1. Association of whistler-mode wave groups with solar wind structures

|  | Coherent (mean; median; mode) | Incoherent (mean; median; mode) |
|---|---|---|
| Temperature anisotropy | 0.92; 0.92; 0.90 | 0.82; 0.83; 0.90 |
| Total heat flux, ergs cm$^{-2}$ s$^{-1}$ | 0.012; 0.010; 0.008 | 0.011; 0.008; 0.00 |
| Total beta | 0.77; 0.70; 0.50 | 0.39; 0.32; 0.25 |
| Solar wind speed, km s$^{-1}$ | 427; 402; 385 | 440; 394; 326 |

Table 2. The mean, median and mode values of electron parameters and solar wind speed for the set of narrowband (coherent) and incoherent whistler mode waves.

ACKNOWLEDGEMENTS: We thank the STEREO PLASTIC Investigation (A.B. Galvin, PI) and NASA grant NNX15AU01G, the IMPACT investigation (J.Luhmann,PI), and B. Lavraud and the SWEA team. The work at the University of Minnesota was supported by NASA grants NNX16AF80G, 80NSSC19K305 and NNX14AK73G. SPEDAS software (courtesy Space Science Laboratory, UC Berkeley) was utilized.

REFERENCES

Bale, S. D., Pulupa, M., Salem, C., Chen, C. H. K., & Quataert, E. 2013, ApJL, 769, L22

Bošková, J., Tříska, P., Omelchenko, Y.A. *et al.* 1992, *Stud Geophys Geod* **36**, 177–187

Bougeret, J.L., et al., 2008. SSRv 136, 487–528.

Breneman, A., Cattell, C., Schreiner, S., Kersten, K., Wilson, L.B., III, Kellogg, P., Goetz, K., Jian, L.K., 2010, JGRA., 115, A08104.

Cattell, C., et al., 2008. GeoRL, 35, L01105.

Cohen, Z. A., C. A. Cattell, A. W. Breneman, L. A. Davis, P. Grul, K. Kersten, L. B. Wilson III, and J. R. Wygant 2019, https://arxiv.org/abs/1909.08176

Coroniti, F. V., C. F. Kennel, F. L. Scarf, and E. J. Smith 1982, JGR, 87, 6029–6044

Ergun , R., et al., 1998, ApJ, 503, 435

Feldman W. C., Asbridge J. R., Bame S. J., Montgomery M. D., Gary S. P., 1975, JGR, 80, 4181

Forslund, D. W., 1970, *JGR.*, **75**, 17, 1970.

Galvin, A.B., Kistler, L.M., Popecki, M.A., Farrugia, C.J., Simunac, L., Ellis, E., et al. 2008, SSRv, 136, 437 – 486

Gary, S. P., Feldman, W. C., Forslund, D. W., & Montgomery, M. D. 1975, *JGR.*, **80**, 4197

Gary, S. P. 1978, *JGR*, **83**, 2504

Gary, S.P., Scime, E.E., Phillips, J.L., and Feldman, W.C. 1994, JGR, 99, 23391–23399.

Gary, S. P., & Wang, J. 1996, *JGR*, **101**, 10749

Gary, S. P., Neagu, E., Skoug, R. M., & Goldstein, B. E. 1999, JGR, 104, 19843

Graham, G. A., et al. 2017, *JGRA*, 122, 3858– 3874

Gurgiolo, C., Goldstein, M. L., Viñas, A. F., and Fazakerley, A. N. 2012, AnGeo, 30, 163—175

Gurnett, D. A., and Anderson, R. R. 1977, JGR, 82( 4), 632— 650,

Jian, L., https://stereo-ssc.nascom.nasa.gov/pub/ins_data/impact/level3/

Jian, L.K., C.T. Russell, J.G. Luhmann, and A.B. Galvin 2018, ApJ, 885, 114,

Jian, L.K., J.G. Luhmann, C.T. Russell, and A.B. Galvin 2019, SoPh, 294, 31.

Kajdič, P., Alexandrova, O., Maksimovic, M., Lacombe, C., Fazakerley, A.N., 2016. ApJ, 833, 172

Kellogg, P. J., K. Goetz, N. Lin, S. J. Monson, A. Balogh, R. J. Forsyth, R. G. Stone,1992, GeoRL, **19**, 1299, 1992

Kennel, C. F., and Petschek, H. E. 1966, JGR, 71( 1), 1— 28,

Kennel, C.F., Scarf, F.L., Coroniti, F.V., Fredricks, R.W., Gurnett, D.A. and Smith, E.J. 1980, GeoRL, 7: 129-132.

Krafft, C. and Volokitin, A.: 2003, AnGeo, 21, 1393—1403

Kuzichev I. V., Vasko I. Y., Soto-Chavez A. R., Tong Y., Artemyev A. V., Bale S. D., Spitkovsky A., 2019, ApJ, 882, 81

Lacombe, C., Alexandrova, O., Matteini, L., Santolík, O., Cornilleau-Wehrlin, N., Mangeney, A., de Conchy, Y., Maksimovic, M., 2014, ApJ, 796

Lemons, D., & Gary, S. 1976, *JPlPh, 15*(1), 83-89.

Lin, N., Kellogg, P. J., MacDowall, R. J., Scime, E. E., Balogh, A., Forsyth, R. J., McComas, D. J., and Phillips, J. L. 1998, JGRA 103(A6), 12023–12035.

Luhmann, J.D., Curtis, W., Schroeder, P., McCauley, J., Lin, R.P., Larson, D.E., et al., 2008, SSRv 136, 117–184.

Maksimovic, M., et al. 2005, JGRA, 110, A09104

Moullard ,O,. ,D . Burgess, S .D . Bale, 1998,A&A, 335(2), 703-708

Moullard, O., Burgess, D., Salem, C., Mangeney, A., Larson, D.E., Bale, S.D., 2001, JGR., 106, 8301–8314.

Neubauer,F. M., Musmann, G., and Dehmel, G. 1977, JGR, 82(22), 3201–3212

Pistinner, S. L. and D. Eichler, 1998, MNRAS, 301 (1), 49–58

Salem, C., Hubert, D., Lacombe, C., et al. 2003, ApJ, 585, 1147

Sauer, K., and R. D. Sydora 2010, AnGeo, 28, 1317–1325.

Sauvaud, J.A., et al. (2008) In: Russell C.T. (eds) The STEREO Mission. Springer, New York, NY

Scime E. E., Bame S. J., Feldman W. C., Gary S. P., Phillips J. L., Balogh A., 1994, JGRA,99, 23401

Shaaban, S M, M Lazar, S Poedts, 2018,MNRAS, 480 (1), 310–319

Sharma, R.P., Tripathi, Y.K., Al Janabi, A.H., Boswell, R.W., 1992,JGRA, 97, 4275–4281

Stansby, D. et al., 2016, ApJL, 829, L16

Tong, Yuguang *et al* 2019 *ApJ* **878** 41

Vasko T. *et al* 2019 *ApJL* **871** L29

Wilson III, Lynn B. *et al.* 2019, *ApJS* **245** 24